\RequirePackage{fix-cm}
\documentclass[smallextended]{svjour3}       
\smartqed  
\usepackage{graphicx}
\usepackage{epstopdf}
\usepackage{amsmath}
\usepackage{color}
\usepackage{natbib}
\begin{document}

\title{Modelling the income distribution in the European Union: An application for the initial analysis of the recent worldwide financial crisis
}

\titlerunning{Physical modelling of income distribution}        

\author{Maciej Jagielski         \and
        Ryszard Kutner 
}


\institute{Institute of Experimental Physics\\
Faculty of Physics, University of Warsaw\\
Ho\.za 69, PL-00681 Warszawa, Poland
              Tel.: +48-22-5533229\\
              \email{zagielski@interia.pl}           
}

\date{Submitted to Journal of Economic Interaction and Coordination}

\maketitle

\begin{abstract}
By using methods of statistical physics, we focus on the quantitative analysis of the economic income data descending from different databases. To explain our approach, we introduce the necessary theoretical background, the extended Yakovenko et al. (EY) model. This model gives an analytical description of the annual household incomes of all society classes in the European Union (i.e., the low-, medium-, and high-income ones) by a single unified formula based on unified formalism. We show that the EY model is very useful for the analyses of various income datasets, in particular, in the case of a smooth matching of two different datasets. The completed database which we have constructed using this matching emphasises the significance of the high-income society class in the analysis of all household incomes. For instance, the Pareto exponent, which characterises this class, defines the Zipf law having an exponent much lower than the one characterising the medium-income society class. This result makes it possible to clearly distinguish between medium- and high-income society classes. By using our approach, we found that the high-income society class almost disappeared in 2009, which defines this year as the most difficult for the EU. To our surprise, this is a contrast with 2008, considered the first year of a worldwide financial crisis, when the status of the high-income society class was similar to that of 2010. This, perhaps, emphasises that the crisis in the EU was postponed by about one year in comparison with the United States. 
\end{abstract}

\section{Introduction}
\label{intro}
In order to describe the complexity of the world around us, contemporary scientific research quite often combines methodologies and methods that have up to now been used in different, even far-away fields of science \citep{R_2002} and not only those offered separately by various scientific disciplines. A prominent example of such research provides the econophysics. This emerging branch of science applies methods and models used most often in statistical physics and condensed matter physics to describe some economic and financial processes. The term \emph{econophysics} was first used by the physicist H. Eugene Stanley in Ref. \citep{SAABGHLMMPPSSV_1996} -- this paper is a kind of manifesto of econophysics.

The term \emph{econophysics} is a neologism -- a combination of two words: economics and physics, as in the case of astrophysics, geophysics, and biophysics, which use methods of physics to describe the phenomena studied within astronomy, geology, and biology, respectively. It should be emphasised that econophysics does not apply the laws of physics literally to describe the economic behaviour of different types of entities, such as investors, individuals, or households. Most often it uses reinterpreted and properly modified methods developed in statistical physics to analyse the statistical properties of complex systems (consisting of a large number of the aforementioned entities) \citep{YR_2009}. Shortly speaking, the econophysics focuses on the quantitative analysis of economic and financial data by the mathematical and physical modelling of a large number of interacting economic entities (also called \emph{agents}) -- it also has much in common with research in econometrics and multi-agent modelling (called \emph{agent-based modelling}) \citep{YR_2009}.

One of the major trends in econophysics is the study of income and wealth redistribution in society and the analysis of social inequalities. Vilfredo Pareto, Italian economist and sociologist, is a pioneer of this research. At the fall of the nineteenth century, Pareto was the first to provide an analytical description of the distribution of wealth in society represented by the annual income of individuals. One of his most significant findings was the fact that the distribution of income of individuals from different countries is universal and has a small variability in "space" and time. In addition, Pareto stated that these distributions do not resemble the shape of the distribution that one would obtain if the accumulation of wealth was a random process. He also recognized the stability of these distributions. That is, even if one excludes the richest and the poorest from the process of gaining income, after a period of time, the distribution of income will resemble the shape of the initial distribution \citep{P_1897,M_1960,RHCR_2006,mgr}. The striking result of Pareto's study was that the distribution of income in countries with a stable economy is described by a universal power-law known nowadays as the Pareto law. As a possible origin of this law, Pareto pointed to a hierarchical, self-similar structure of societies. Pareto's findings inspired many researchers to continue attempting analytical descriptions of the income of societies.

The income of societies has also been analysed by French economist Robert Gibrat. He pointed out that the Pareto law is not able to describe the distribution of income over the entire range. He proposed a complementary approach known as the Rule of Proportionate Growth. Using the stochastic process to describe the dynamics of income or wealth of a single person or household, he found out that the theoretical probability distribution function described incomes belonging to the low-income society class \citep{G_1931,K_1945,A_1995,S_1997,RHCR_2006,mgr}.

Furthermore, David Champernowne proposed one of the first stochastic models, which reproduces the Pareto law \citep{Ch_1953, mgr}. Benoit Mandelbrot described the fundamental properties of random variables from the Pareto distribution \citep{M_1963, mgr}.

The analytical description of the distribution functions of income made by Gibrat, Champernowne and Mandelbrot led to the disclosure of many significant properties of these distributions, however, it did not give an answer to the crucial question concerning the microscopic (microeconomic) mechanism determining empirical complementary distribution functions. Recently, several models have been proposed which partially explain the microscopic mechanisms (for income dynamics of individuals or households) standing behind the observed empirical complementary distribution functions of incomes. 

In principle, the above mentioned models can be classified into two different groups. The first group is based on Boltzmann's kinetic theory of collision in gases. Analogously to the collision of two particles, if kinetic energy is exchanged between them, it is assumed in models -- called collision models -- that two randomly selected individuals or households exchange money according to the corresponding rules \citep{A_1986, A_1992, A_1993, A_1996, IKR_1998, CC_2000, DY_2000, A_2002, A_2006}. By using more complex money exchange rules we obtain the basic collision model proposed by Dr\u{a}gulescu and Yakovenko \citep{DY_2000, CC_2007}, which leads to the Boltzmann-Gibbs distribution. This group also contains: (i) models of collisions allowing a negative income (or debt) \citep{DY_2000, FB_2003, XDW_2005, CC_2008}, (ii) models of collisions with the same saving propensity for all individuals \citep{A_1986, A_1992, A_1993, A_1996, IKR_1998, A_2002, PCK_2004, CC_2007}, and (iii) models of collisions with varying saving propensity \citep{CC_2000, A_2002, PCK_2004a, CCM_2004, F_2004, SPW_2004a, SPW_2004b, PCKG_2005, RHR_2005, Ch_2005, CCS_2005, BCC_2005, A_2006, CC_2007}. The second group of models treats the income of an individual or household as a random variable. To describe the dynamics of this variable, the nonlinear stochastic Langevin equation and the corresponding Fokker--Planck equation are used. Depending on the specific assumptions concerning the dynamics of income, one can obtain the following models: (i) the Boltzmann--Gibbs law \citep{RHCR_2006, YR_2009, BY_2010}, (ii) the Pareto law \citep{RHCR_2006, YR_2009, BY_2010}, (iii) the Rule of Proportionate Growth \citep{G_1931, K_1945, A_1995, S_1997, RHCR_2006, YR_2009}, (iv) the Generalised Lotka--Volterra model \citep{SR_2001, RS_2001, SR_2002, H_2004, RHCR_2006, YR_2009}, and (v) the Yakovenko et al. model \citep{YR_2009, BY_2010}. Remarkably, both in the case of the Boltzmann kinetic theory and in the case of Langevin stochastic dynamics (which lead to distributions of income of individuals or households), econophysics paves the way for new trends of research, complementary to those developed in economic and social sciences \citep{KW_1993, M_2006}.

Besides the construction of analytical models describing the distribution functions of income, their verifications were intensively conducted recently, as extensive empirical databases became, in principle, a public domain. These verifications were carried out, among others, for the United States \citep{LS_1997, DY_2001a, DY_2001b, R_2003, RRR_2004, SPW_2004a, LO_2004, YS_2005, SY_2005, CG_2005a, NS_2007, CMGK_2008, CGK_2009, YR_2009, BY_2010}, the United Kingdom \citep{P_1897, DY_2001a, SPW_2004a, SPW_2004b, F_2004, F_2005, CG_2005a, RHCR_2006, CGK_2007}, Germany \citep{P_1897, CG_2005a, CGK_2007}, Italy \citep{P_1897, CG_2005b, CMG_2006, CGK_2007}, France \citep{P_1897}, Switzerland \citep{P_1897}, Japan \citep{ASNOTT_2000, S_2001, FSAKA_2003, ASF_2003, F_2004, F_2005, SN_2005, NS_2007}, Australia \citep{MAH_2004, CMG_2006, BYM_2006, CMGK_2008}, Canada \citep{R_2003}, the Czech Republic \citep{R_2003}, New Zealand \citep{F_2004, F_2005}, India \citep{S_2006}, Sri Lanka \citep{R_2003}, Argentina \citep{F_2005}, Peru \citep{P_1897}, South Korea \citep{KY_2004}, and Romania \citep{DNS_2012}.

However, none of the models that have been developed so far (to the best of our knowledge) give an analytical description of the annual household incomes of all society classes (i.e. the low-, medium-, and high-income society classes) by a single unified formula based on unified formalism. In our recent papers \citep{JK_2013a, JK_2013b} we developed the Extended Yakovenko model, which provided such a powerful formula. This formula (with the low number of free parameters) reproduces the empirical complementary cumulative distribution function in the entire range of the income. In the present paper we give a short review of the Extended Yakovenko model and show that this model is valid for various income datasets, especially when matching two different datasets.  

It should be noted that the subject of this paper relates, at least partially, to the problems analysed in sociophyics. Sociophyics, in contrast to econophysics, does not focus solely on the research of economic activity of individuals but, by using the methods of physics, it studies the social mainstream subjects such as the analysis of political preferences, social networks, formed coalitions, terrorism as well as the dynamics of public opinions and emotions \citep{YR_2009, G_2012}. 

\section{Matched dataset}\label{section:MDS}

We exploit the empirical data from Eurostat's Survey on Income and Living Conditions (EU--SILC) \citep{EURO, EURO_2005, EURO_2006, EURO_2007, EURO_2008, EURO_2009, EURO_2010} for the years 2005--2010. This database contains information on the demographic characteristics of households in the European Union (EU), their living conditions, and their income and economic activity. For our analysis we chose the \emph{Total household gross income} variable. According to Eurostat, the definition of the annual total gross household income (we quote) "\ldots is the total monetary and non-monetary income of a household over a period of one year, before deducting taxes on income or wealth or social security contributions by employers and employees but after including inter-household transfers received" \citep{EURO}.

Eurostat's EU--SILC database contains only a few records concerning the income of households belonging to the high-income society class. That is, these households cannot be subject to any statistical description. Therefore, in order to consider the high-income society class, we have additionally analysed the effective income of billionaires in the EU by using the Forbes ranking "The World's Billionaires"\footnote{The term 'billionaire' used herein is equivalent (as in US terminology) to the term 'multimillionaire' used in European terminology. Since we consider the wealth and income of billionaires in euros, we recalculated the US dollars to euros by using the average  exchange rate on the day the Forbes list "The World's Billionaires" was constructed.}${}^{,}$\footnote{The billionaires who gained effective incomes are billionaires whose incomes are greater than zero.} \citep{FORBES}. This ranking contains individuals whose value of wealth in a given year exceeds one billion US dollars. From this ranking, we selected only those who reside in the European Union. Hence, we were able to make the ranking of the richest Europeans for the years 2004--2010.

Next, by using the EU--SILC database and the ranking of the richest Europeans, we considered incomes of three society classes thanks to the following procedure (roughly described in Refs. \citep{JK_2013a, JK_2013b}):
\begin{itemize}
\item[(i)] Firstly, we calculated their incomes for the years 2005--2010. This calculation was possible because we assumed that the billionaires' incomes were proportional to the corresponding differences between their wealth for the pairs of successive years (here from 2004 to 2010). Notably, we took into account only the billionaires who gained effective incomes.
\item[(ii)] Secondly, having calculated the incomes for the high-income society class, we simply matched them with the EU--SILC dataset. Then, by using the dataset completed in this way, we constructed the initial empirical complementary cumulative distribution function for the years 2005--2010 separately. For that, we used the well known Weibull recipe (see below for details) \citep{H_2004, CMM_1988}. However, this direct but too simplified approach shows a wide gap of incomes among the high-income society class resulting in a horizontal line of the complementary cumulative distribution function. 
The reason for the gap is that the first segment of high-income society class consists in all the data points derived from the EU--SILC dataset, whereas the other segment of high-income society class, comprised of the remaining data points, has been taken from the Forbes dataset.
\item[(iii)] In the final step, we eliminated this gap by adopting the assumption that the empirical complementary cumulative distribution function
(concerning the whole society) has no horizontal segments. That is, we assumed that the statistics of income is a continuous function of income (i.e. there is no disruption). Hence, we were forced to multiply the billionaire incomes from the Forbes dataset by the properly chosen common proportionality factor. This factor is not an arbitrary one -- it is equal to $1.0\times 10^{-2}$, since the obvious requirement of a full overlap of the first (above mentioned) segment by the subsequent (second) segment was assumed. Hence, this approach leads to a unique solution (up to some negligible statistical error) for this proportionality factor. Furthermore, we found that this factor was only a slowly-varying function of time (or years).
\end{itemize}

Thus, we obtained the matched dataset (MDS) already containing the sufficient data points covering all society classes, i.e. containing also the high-income society class. 

In order to analyse the presented empirical data in a more stable form, we used an empirical complementary cumulative distribution function. We calculated it according to the commonly used two-step procedure. Firstly, the income empirical data was ordered according to its rank, i.e. from the richest household incomes to the poorest. Next, in accordance with the Weibull formula \citep{CMM_1988, Han_2004}, we calculated the ratio $\frac{l}{n+1}$, where $l$ is the position of the household in the rank and $n$ is the size of the empirical data record. This ratio directly determines the required fraction of households of an income higher than that related to a given household position $l$ in the rank. The complementary cumulative distribution function obtained this way is sufficiently persistent. Furthermore, it does not reduce the size of the output in comparison to that of the original empirical data record.

\section{Extended Yakovenko et al. model}\label{section:eYm}

We present the necessary theoretical background that is extended by us, the Yakovenko et al. model. A detailed description of the model can be found in our earlier paper \citep{JK_2013b}.

Let $m$ be an influx of income per unit of time for a given household. We treat $m$ as a variable obeying the stochastic dynamics. Then, we can describe the time evolution of the income probability distribution function by using the so called second diffusion equation\footnote{Physicists call the second diffusion equation the Fokker--Planck equation. This equation is equivalent to the Langevin equation
\begin{eqnarray}
\frac{dm}{dt}=-A(m)+C(m)\, \eta(t).
\label{rown1}
\end{eqnarray}
Here, $A(m)$ is a drift term and $\eta(t)$ is a white noise, where the coefficient $C(m)$ is its $m$-dependent amplitude.} \citep{YR_2009, BY_2010, vanK_1990}
\begin{eqnarray}
\frac{\partial}{\partial t}P(m,t)&=&\frac{\partial}{\partial m}[A(m)P(m,t)]+
\frac{{\partial ^2}}{\partial m^2}\left[B(m)P(m,t)\right]. 
\label{rown2a}
\end{eqnarray}
where, $B(m)=C^2(m)/2$ and $P(m,t)$ is the temporal income distribution function. In general, functions $A(m)$ and $B(m)$ can additionally be determined by the first and second moments of the income change per unit of time, respectively, only if these moments exist. Subsequently, the equilibrium solution of Eq. (\ref{rown2a}), $P_{\rm eq}$ takes the form \citep{vanK_1990}:
\begin{eqnarray}
P_{\rm eq}(m)=\frac{const}{B(m)}\exp\left(-\int_{{m_{\rm init}}}^m\frac{A(m')}{B(m')}\, dm'\right)
\label{rown5}
\end{eqnarray}
where $m_{\rm init}$ is the lowest household income and $const$ is a normalisation factor. 

Using Eq. (\ref{rown5}) we derive such a distribution function which covers all three ranges of the empirical data records, i.e. the low-, medium-, and high-income society classes (including also two short intermediate regions between them). Therefore, we provide function $A(m)$ in a threshold form \citep{JK_2013a, JK_2013b}: 
\begin{eqnarray}
A(m) &=& \left\{ \begin{array}{ll}
A^<(m)=A_0+a\, m & \textrm{if $m<m_1$} \\
{A^{\ge }(m)=A'_0}+a'\, m & \textrm{if $m\ge m_1$,}
\label{rown12}
\end{array} \right. \nonumber\\
B(m)&=&B_0+b\, m^2=b\, (m^2_0+m^2),
\end{eqnarray}
where parameters used in the above equation are defined and considered below.

This definition of $A(m)$ and $B(m)$ allows for the coexistence of additive and multiplicative stochastic processes since we assume that household income consists of two components. The first, a pure deterministic component of income, arises from the fact that household income is determined by wages and salaries. The second, already an indeterministic component, expresses profits which go to households mainly through investments and capital gains.

The threshold parameter $m_0$ is the crossover income between the low- and medium-income society classes, while parameter $m_1$ is interpreted as a crossover income between the medium- and high-income society classes. 

Subsequently, by substituting Eq. (\ref{rown12}) into Eq. (\ref{rown5}), we finally get
\begin{eqnarray}
P_{\rm eq}(m) =\left\{ \begin{array}{ll}
c'\, \frac{\exp\left(-(m_0/T)\arctan(m/m_0)\right)}{[1+(m/m_0)^2]^{(\alpha +1)/2}}, & \textrm{if $m<m_1$} \\
c''\, \frac{\exp\left(-(m_0/T_1)\arctan(m/m_0)\right)}{[1+(m/m_0)^2]^{(\alpha_1 +1)/2}}, & \textrm{if $m\ge m_1$}
\end{array} \right. \label{rown19}
\end{eqnarray}
where exponents $\alpha=1+a/b$, $\alpha _1=1+a'/b$, and income temperatures $T=B_0/A_0$, $T_1=B_0/A'_0$. Parameter $T$ can be interpreted in this case as the average income per household for low- and medium-income society classes. Parameter $T_1$ has the same interpretation but for a high-income society class. Apparently, the number of free (effective) parameters driving the two-branch distribution function, given by Eq. (\ref{rown19}), is reduced because this function depends only on the ratios of the corresponding parameters defining the nonlinear Langevin dynamics given by Eq. (\ref{rown1}).  

For $m\ll m_0$ the distribution function, given by the first expression in Eq. (\ref{rown19}), becomes the Boltzmann--Gibbs law. For $m_0\ll m<m_1$ it gives the weak Pareto law with exponent $\alpha$. For $m \gg m_1$ we obtain, from the second expression in Eq. (\ref{rown19}), also the weak Pareto law but with the exponent $\alpha_1<\alpha$.

\section{Results and concluding remarks}

In this Section we compared the theoretical complementary cumulative distribution functions, based on Eq. (\ref{rown19}), with the corresponding empirical ones. The latter are constructed from data coming from two sources -- the EU--SILC dataset and the MDS.  

\begin{figure}[ht]
\centering
\includegraphics[scale=0.50]{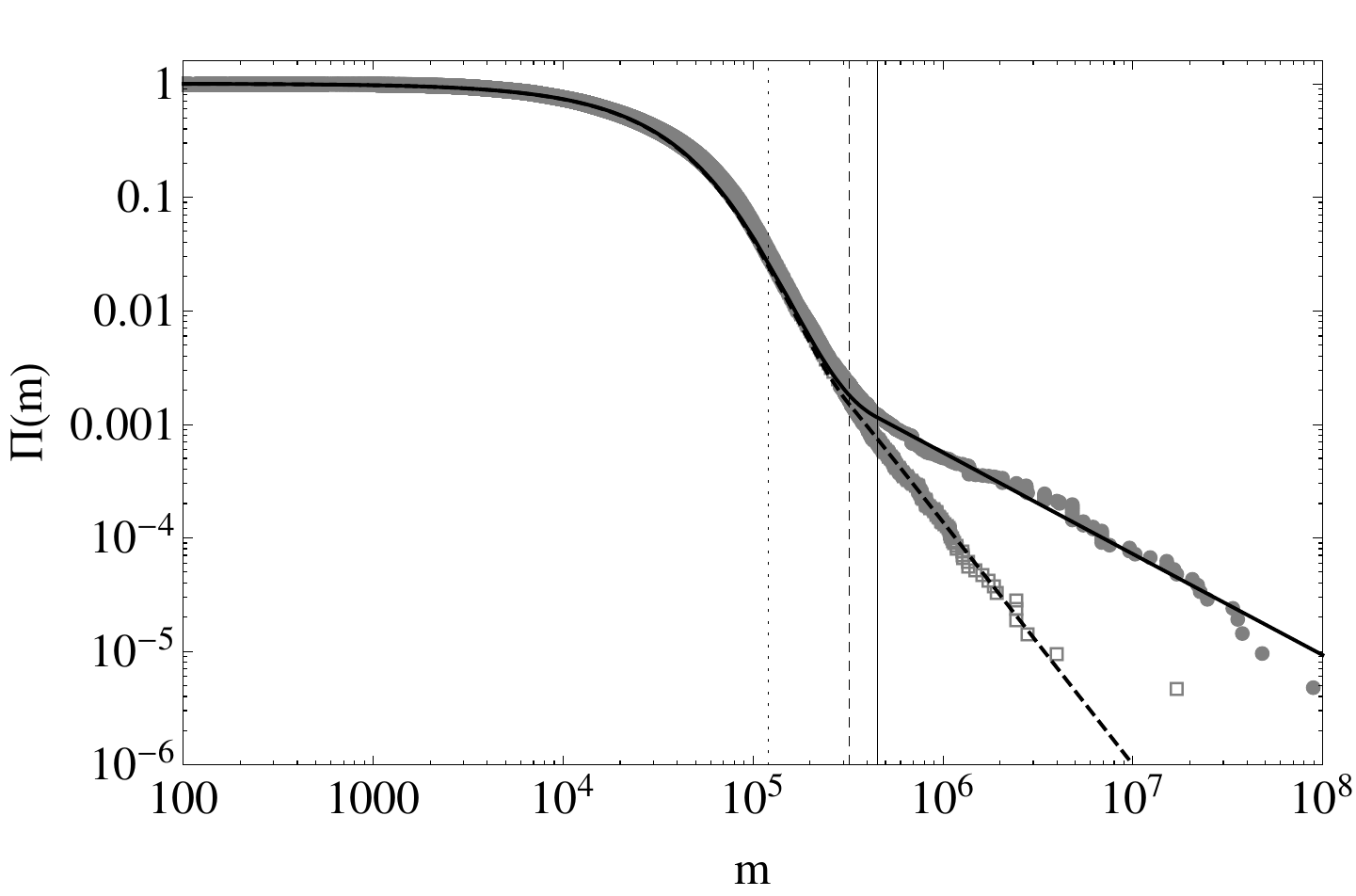}
\caption{Two plots of complementary cumulative distribution functions obtained from the extended Yakovenko et al. formula (\ref{rown19}) (dashed and solid curves) and income empirical datasets (open squares and full circles; cf. \citep{EURO_2008, FORBES} for details). The lower plot (dashed curve and open squares) concerns the EU--SILC database. The upper plot (solid curve and full circles) concerns MDS -- both for the year 2008. The dotted vertical line denotes the value $m_0$ common for both plots. The dashed and solid vertical lines denote two different values of $m_1$ for both plots, respectively.}
\label{fig1}
\end{figure}

\begin{figure}[ht]
\centering
\includegraphics[scale=0.50]{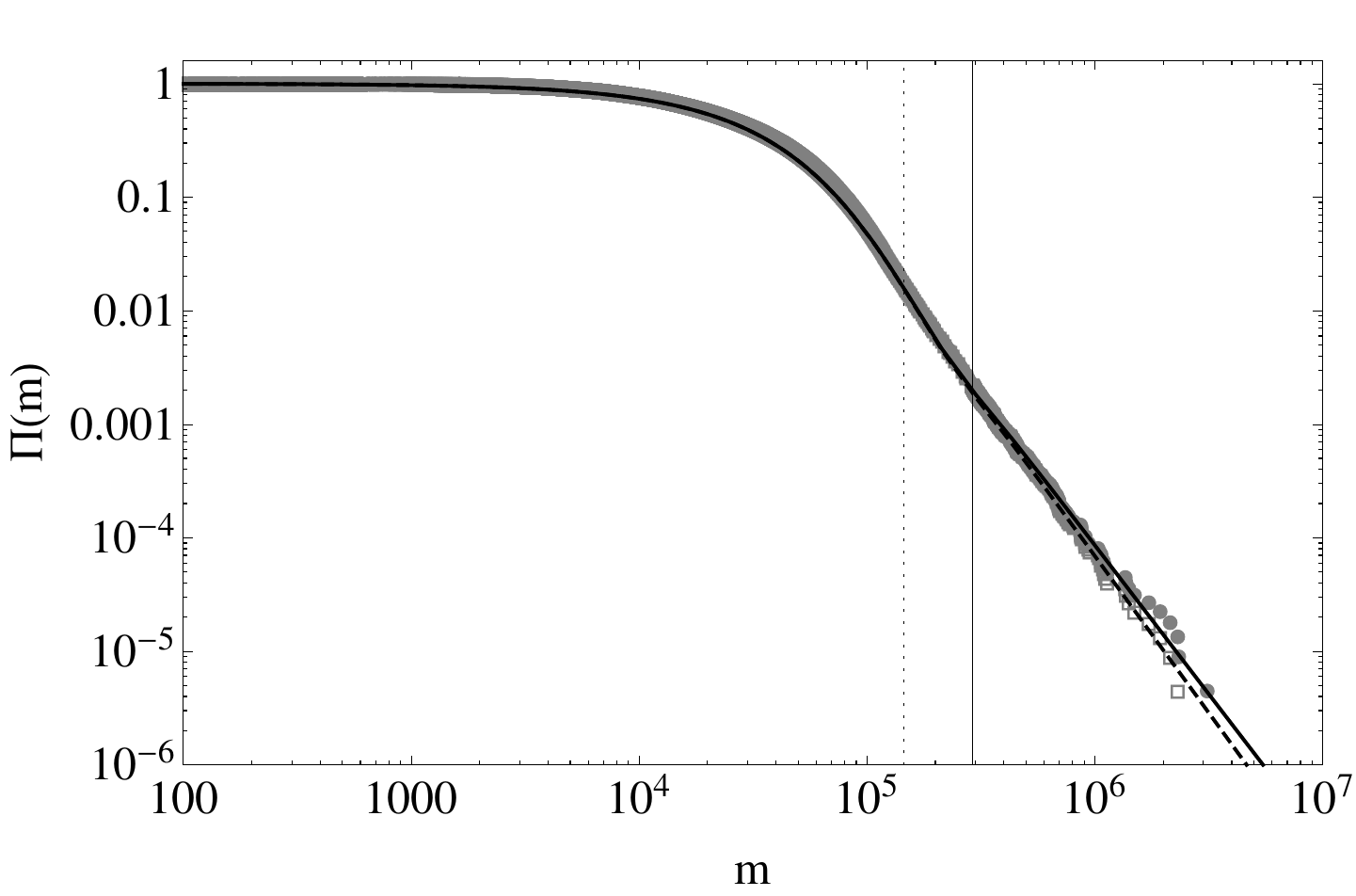}
\caption{Two, almost overlapping plots of complementary cumulative distribution functions obtained from the EY Formula (\ref{rown19}) (both dashed and solid curves) and income empirical datasets (open squares and full circles as well; cf. \citep{EURO_2009, FORBES} for details). The slightly lower plot (dashed curve and open squares) concerns the EU--SILC database. The slightly upper plot (solid curve and full circles) concerns MDS -- both for the year 2009. The dotted vertical line denotes the value $m_0$ common for both plots. The solid vertical line denotes the value $m_1$, common for both plots.}
\label{fig2}
\end{figure}

\begin{figure}[ht]
\centering
\includegraphics[scale=0.50]{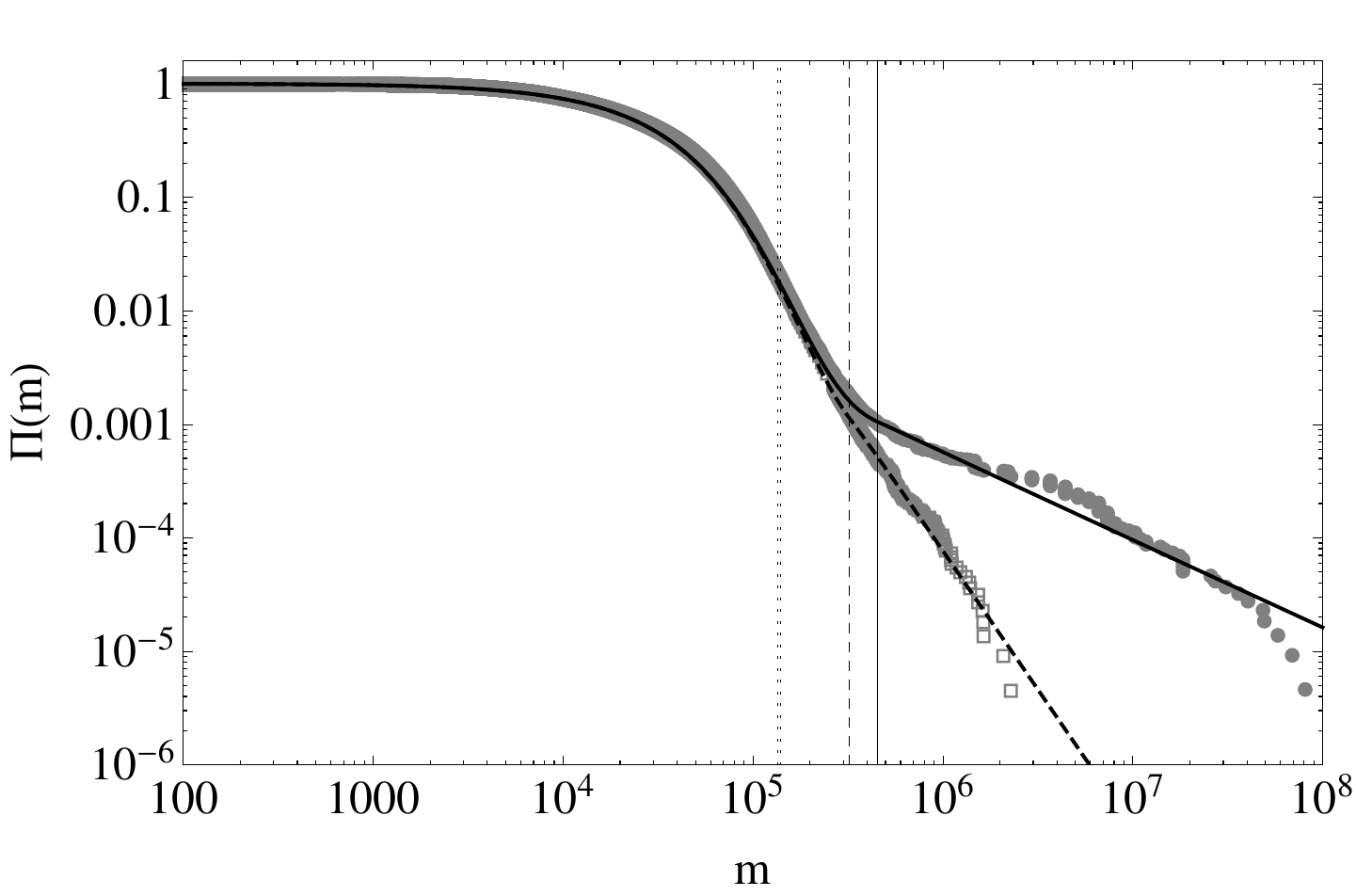}
\caption{Two plots of complementary cumulative distribution functions (very similar to those shown in Fig.\ref{fig1}) obtained from Formula (\ref{rown19}) (both dashed and solid curves) and income empirical datasets (open squares and full circles; cf. \citep{EURO_2010, FORBES} for details). The lower plot (dashed curve and open squares) concerns the EU--SILC database. The upper plot (solid curve and full circles) concerns MDS -- both for the year 2010. The dotted vertical line denotes the value $m_0$ common for both plots. The dashed and solid vertical lines denote two different values of $m_1$ for both plots, respectively.}
\label{fig3}
\end{figure}

The corresponding plots of the theoretical and empirical complementary cumulative distribution functions are compared, in a log-log scale, in Figs. \ref{fig1}--\ref{fig3} for the most intriguing years 2008--2010, respectively. Besides, for further comparisons, Tables \ref{tab1}--\ref{tab3} provide estimates of the parameters of the EY formula for the years 2005--2010 for both above mentioned datasets.

Notably, the high-income society class almost disappeared in 2009, which can define this year as the economic crash for the EU. This is in contrast to 2008, which was defined as a worldwide financial crisis, when the status of the high-income society class was very similar to that of 2010. This, perhaps, emphasises that the crisis in the EU was postponed by about one year in comparison with that of the United States. This is the striking utilitarian result of the paper.

\begin{table}[ht]
\centering
\caption{Values of parameters $T$ and $T_1$ found by the fit of the EY Formula (\ref{rown19}) as well: (i) to the empirical cumulative distribution functions of the annual total gross income of households obtained from the EU--SILC database as (ii) from the EU--SILC+Forbes database -- the years 2005--2010.}
 \begin{tabular}{|c|c|c||c|c|c|c|c|}
  \hline
 & \multicolumn{2}{|c||}{ \bf EU--SILC} & \multicolumn{2}{|c|}{ \bf EU--SILC+Forbes}\\
 & \multicolumn{2}{|c||}{ \bf database} & \multicolumn{2}{|c|}{ \bf database}\\
  \hline 
 {\bf Year} & $\boldsymbol{T}$ {\bf [EUR]} & $\boldsymbol{T_1}$ {\bf [EUR]} & $\boldsymbol{T}$ {\bf [EUR]} & $\boldsymbol{T_1}$ {\bf [EUR]}\\
\hline
	2005 & $36\,000\pm 3000$ & $390\,000\pm 50\,000$ & $36\,000\pm 3000$ & $430\,000\pm 50\,000$ \\
  \hline
	2006 & $37\,000\pm 3000$ & $330\,000\pm 50\,000$ & $37\,000\pm 3000$ & $445\,000\pm 50\,000$\\
  \hline
	2007 & $37\,000\pm 3000$ & $325\,000\pm 50\,000$ & $37\,000\pm 3000$ & $480\,000\pm 50\,000$\\
  \hline
	2008 & $38\,000\pm 3000$ & $320\,000\pm 50\,000$ & $38\,000\pm 3000$ & $450\,000\pm 50\,000$\\
  \hline
	2009 & $37\,000\pm 3000$ & $290\,000\pm 50\,000$ & $37\,000\pm 3000$ & $290\,000\pm 50\,000$\\
  \hline
	2010 & $38\,000\pm 3000$ & $320\,000\pm 50\,000$ & $38\,000\pm 3000$ & $450\,000\pm 50\,000$\\
	\hline
\end{tabular} 
\label{tab1}
\end{table}

\begin{table}[ht]
\centering
\caption{Values of parameters $m_0$ and $m_1$ found by the fit of the EY Formula (\ref{rown19}) as well: (i) to the empirical cumulative distribution functions of the annual total gross income of households obtained from the EU--SILC database as (ii) from the EU--SILC+Forbes database -- the years 2005--2010.}
 \begin{tabular}{|c|c|c||c|c|c|c|c|}
  \hline
 & \multicolumn{2}{|c||}{ \bf EU--SILC} & \multicolumn{2}{|c|}{ \bf EU--SILC+Forbes}\\
 & \multicolumn{2}{|c||}{ \bf database} & \multicolumn{2}{|c|}{ \bf database}\\
  \hline 
 {\bf Year} & $\boldsymbol{m_0}$ {\bf [EUR]} & $\boldsymbol{m_1}$  {\bf [EUR]} & $\boldsymbol{m_0}$  {\bf [EUR]} & $\boldsymbol{m_1}$  {\bf [EUR]}\\
\hline
	2005 & $160\,000\pm 20\,000$ & $390\,000\pm 50\,000$ & $155\,000\pm 20\,000$ & $430\,000\pm 50\,000$ \\
  \hline
	2006 & $150\,000\pm 20\,000$ & $330\,000\pm 50\,000$ & $145\,000\pm 20\,000$ & $445\,000\pm 50\,000$\\
  \hline
	2007 & $160\,000\pm 20\,000$ & $325\,000\pm 50\,000$ & $160\,000\pm 20\,000$ & $480\,000\pm 50\,000$\\
  \hline
	2008 & $120\,000\pm 20\,000$ & $320\,000\pm 50\,000$ & $120\,000\pm 20\,000$ & $450\,000\pm 50\,000$\\
  \hline
	2009 & $145\,000\pm 20\,000$ & $290\,000\pm 50\,000$ & $145\,000\pm 20\,000$ & $290\,000\pm 50\,000$\\
  \hline
	2010 & $140\,000\pm 20\,000$ & $320\,000\pm 50\,000$ & $135\,000\pm 20\,000$ & $450\,000\pm 50\,000$\\
	\hline
\end{tabular} 
\label{tab2}
\end{table}

\begin{table}[ht]
\centering
\centering
\caption{Values of exponents $\alpha $ and $\alpha_1$ found by the fit of the EY Formula (\ref{rown19}) as well: (i) to the empirical cumulative distribution functions of the annual total gross income of households obtained from the EU--SILC database as (ii) from the EU--SILC+Forbes database -- the years 2005--2010.}
 \begin{tabular}{|c|c|c||c|c|c|c|c|}
  \hline
 & \multicolumn{2}{|c||}{ \bf EU--SILC} & \multicolumn{2}{|c|}{ \bf EU--SILC+Forbes}\\
 & \multicolumn{2}{|c||}{ \bf database} & \multicolumn{2}{|c|}{ \bf database}\\
 \hline
 {\bf Year} & $\boldsymbol{\alpha}$ & $\boldsymbol{\alpha_1}$ & $\boldsymbol{\alpha}$ & $\boldsymbol{\alpha_1}$\\
\hline
	2005 & $3.216\pm 0.002$ & $1.54\pm 0.02$ & $2.907\pm 0.003$ & $0.795\pm 0.009$ \\
  \hline
	2006 & $3.094\pm 0.003$ & $2.15\pm 0.02$ & $2.892\pm 0.004$ & $0.86\pm 0.01$\\
  \hline
	2007 & $3.057\pm 0.003$ & $2.32\pm 0.01$ & $2.735\pm 0.004$ & $0.79\pm 0.01$\\
  \hline
	2008 & $3.0632\pm 0.0005$ & $2.13\pm 0.02$ & $2.965\pm 0.001$ & $0.890\pm 0.007$\\
  \hline
	2009 & $2.979\pm 0.001$ & $2.750\pm 0.005$ & $2.974\pm 0.001$ & $2.608\pm 0.006$\\
  \hline
	2010 & $3.329\pm 0.001$ & $2.43\pm 0.01$ & $3.153\pm 0.002$ & $0.77\pm 0.01$\\
	\hline
\end{tabular} 
\label{tab3}
\end{table}

Apparently, the predictions of the Extended Yakovenko et al. Formula (\ref{rown19}), agree with the corresponding empirical cumulative distribution functions of the annual total gross incomes of households in the European Union. We confirmed that the value of parameter $m_0$ can be considered a crossover income between low- and medium-income society classes. Similarly, the value of parameter $m_1$ can be considered as a (subsequent) crossover income between medium- and high-income society classes. The values of exponents $\alpha$ and $\alpha_1$ indicate social stratifications within the medium- and high-income society classes, respectively. The lower values of $\alpha$ and $\alpha_1$ mean higher social stratifications in the corresponding classes. For values of parameters $T$ and $T_1$, we obtained quite reasonable quantities, which confirm their interpretation given in Sec \ref{section:eYm}. 

Furthermore, Tables \ref{tab1}--\ref{tab3} show that regardless of whether the Extended Yakovenko model et al. is compared with empirical data coming from the EU--SILC database or with empirical data coming from the matched databases of EU--SILC and Forbes magazine, MDS, the parameters $T$, $m_0$ and $\alpha$ are surprisingly stable, that is, they differ only slightly for each considered year. Thus, the shapes of the empirical cumulative distribution functions describing the low- and medium-income society classes in the EU, in both cases, are very similar (cf. Figs. \ref{fig1}--\ref{fig3}). On the contrary, we can observe that after matching the EU--SILC with the Forbes database the exponent $\alpha_1$ (describing the tail of the complementary cumulative distribution function, i.e., the high-income society class) strongly decreases (except for the year 2009 -- see Table \ref{tab3} for details). Besides, only a slight extension of the range of the empirical complementary cumulative distribution function referring to the middle-income society class (i.e. the income values between $m_0$ and $m_1$), is observed.

In conclusion, the completed database MDS, which we constructed, emphasises the decisive role of the high-income society class in a thorough, systematic analysis of the annual household incomes in the EU.  

As the extended Yakovenko et al. model describes well the income of the EU households  while the (usual) Yakovenko model \citep{YR_2009, BY_2010} is valid for the incomes of US households, we believe it would be interesting to see the results of a comparative analysis of the incomes of the  EU and the US households.

We suppose that our results give the basis for a better understanding of the mechanisms of enrichment and impoverishment of societies. It is also very likely that we can find a quite precise classification of income ranges which determine whether a given household belongs to the low-, medium- or high-income society class.

\bibliographystyle{spbasic}
\bibliography{bibliography}

\begin{thebibliography}{84}
\providecommand{\natexlab}[1]{#1}
\providecommand{\url}[1]{{#1}}
\providecommand{\urlprefix}{URL }
\expandafter\ifx\csname urlstyle\endcsname\relax
  \providecommand{\doi}[1]{DOI~\discretionary{}{}{}#1}\else
  \providecommand{\doi}{DOI~\discretionary{}{}{}\begingroup
  \urlstyle{rm}\Url}\fi
\providecommand{\eprint}[2][]{\url{#2}}

\bibitem[{Angle(1986)}]{A_1986}
Angle J (1986) The surplus theory of social stratification and the size
  distribution of~personal wealth. Soc Forces 65:293--326

\bibitem[{Angle(1992)}]{A_1992}
Angle J (1992) The inequality process and the distribution of income to blacks
  and whites. J Math Sociol 17:77--98

\bibitem[{Angle(1993)}]{A_1993}
Angle J (1993) Deriving the size distribution of personal wealth from ''the
  rich get richer, the poor get poorer''. J Math Sociol 18:27--46

\bibitem[{Angle(1996)}]{A_1996}
Angle J (1996) How the gamma law of income distribution appears invariant under
  aggregation. J Math Sociol 21:325--358

\bibitem[{Angle(2002)}]{A_2002}
Angle J (2002) The statistical signature of pervasive competition on wage and
  salary incomes. J Math Sociol 26:217--270

\bibitem[{Angle(2006)}]{A_2006}
Angle J (2006) The inequality process as a wealth maximizing process. Physica A
  367:388--414

\bibitem[{Aoyama et~al(2000)Aoyama, Souma, Nagahara, Okazaki, Takayasu, and
  Takayasu}]{ASNOTT_2000}
Aoyama H, Souma W, Nagahara Y, Okazaki M, Takayasu H, Takayasu M (2000)
  Pareto's law for income of individuals and debt of bankrupt companies.
  Fractals 8:293--300

\bibitem[{Aoyama et~al(2003)Aoyama, Souma, and Fujiwara}]{ASF_2003}
Aoyama H, Souma W, Fujiwara Y (2003) Growth and fluctuations of personal and
  company's income. Physica A 324:352--358

\bibitem[{Armatte(1995)}]{A_1995}
Armatte M (1995) Robert {G}ibrat et la loi de l'effet proportionnel.
  Math\'{e}matiques et sciences humaines 129:5--35

\bibitem[{Banerjee and Yakovenko(2010)}]{BY_2010}
Banerjee A, Yakovenko V (2010) Universal patterns of inequality. New J Phys
  12:075{}032

\bibitem[{Banerjee et~al(2006)Banerjee, Yakovenko, and Matteo}]{BYM_2006}
Banerjee A, Yakovenko V, Matteo TD (2006) A study of the personal income
  distribution in {A}ustralia. Physica A 370:54--59

\bibitem[{Bhattacharyya et~al(2005)Bhattacharyya, Chatterjee, and
  Chakrabarti}]{BCC_2005}
Bhattacharyya P, Chatterjee A, Chakrabarti BK (2005) A common mode of origin of
  power laws in models of market and earthquake. Physica A 381:377--382

\bibitem[{Chakrabarti(2005)}]{Ch_2005}
Chakrabarti B (2005) Econophys--kolkata: A short story. In: Chatterjee A,
  Chakrabarti B, Yarlagadda S (eds) Econophysics of Wealth Distributions,
  Springer, Italy, pp 225--228

\bibitem[{Chakraborti and Chakrabarti(2000)}]{CC_2000}
Chakraborti A, Chakrabarti B (2000) Statistical mechanics of money: how saving
  propensity affects its distribution. Eur Phys J B 17:167--170

\bibitem[{Champernowne(1953)}]{Ch_1953}
Champernowne D (1953) A model of income distribution. Econ J 63:318--351

\bibitem[{Chatterjee and Chakrabarti(2007)}]{CC_2007}
Chatterjee A, Chakrabarti B (2007) Kinetic exchange models for income and
  wealth distributions. Eur Phys J B 60:135--149

\bibitem[{Chatterjee et~al(2004)Chatterjee, Chakrabarti, and Manna}]{CCM_2004}
Chatterjee A, Chakrabarti B, Manna S (2004) Pareto law in a kinetic model of
  market with random saving propensity. Physica A 335:155--163

\bibitem[{Chatterjee et~al(2005)Chatterjee, Chakrabarti, and
  Stinchcombe}]{CCS_2005}
Chatterjee A, Chakrabarti B, Stinchcombe RB (2005) Master equation for a
  kinetic model of a trading market and its analytic solution. Phys Rev E
  72:026{}126

\bibitem[{Chow et~al(1998)Chow, Maidment, and Mays}]{CMM_1988}
Chow V, Maidment D, Mays L (1998) Applied Hydrology. McGraw-Hill, Singapore

\bibitem[{Clementi and Gallegati(2005{\natexlab{a}})}]{CG_2005a}
Clementi F, Gallegati M (2005{\natexlab{a}}) Pareto's law of income
  distribution: Evidence for {G}ermany, the {U}nited {K}ingdom, and the
  {U}nited {S}tates. In: Chatterjee A, Chakrabarti B, Yarlagadda S (eds)
  Econophysics of Wealth Distributions, Springer, Italy, pp 3--14

\bibitem[{Clementi and Gallegati(2005{\natexlab{b}})}]{CG_2005b}
Clementi F, Gallegati M (2005{\natexlab{b}}) Power law tails in the italian
  personal income distribution. Physica A 350:427--438

\bibitem[{Clementi et~al(2006)Clementi, Matteo, and Gallegati}]{CMG_2006}
Clementi F, Matteo TD, Gallegati M (2006) The power-law tail exponent of income
  distributions. Physica A 370:49--53

\bibitem[{Clementi et~al(2007)Clementi, Gallegati, and Kaniadakis}]{CGK_2007}
Clementi F, Gallegati M, Kaniadakis G (2007) $\kappa$-generalized statistics in
  personal income distribution. Eur Phys J B 57:187--193

\bibitem[{Clementi et~al(2008)Clementi, Matteo, Gallegati, and
  Kaniadakis}]{CMGK_2008}
Clementi F, Matteo TD, Gallegati M, Kaniadakis G (2008) The
  $\kappa$-generalized distribution: A new descriptive model for the size
  distribution of incomes. Physica A 387:3201--3208

\bibitem[{Clementi et~al(2009)Clementi, Gallegati, and Kaniadakis}]{CGK_2009}
Clementi F, Gallegati M, Kaniadakis G (2009) A $\kappa$-generalized statistical
  mechanics approach to income analysis. Journal of Statistical Mechanics p
  P02037

\bibitem[{Cockshott and Cottrell(2008)}]{CC_2008}
Cockshott P, Cottrell A (2008) Probabilistic political economy and endogenous
  money, unpublished paper, available at:
  http://www.dcs.gla.ac.uk/publications/PAPERS/8935/probpolecon.pdf

\bibitem[{Derzsy et~al(2012)Derzsy, N{\'e}da, and Santos}]{DNS_2012}
Derzsy N, N{\'e}da Z, Santos M (2012) Income distribution patterns from a
  complete social security database. Physica A 391:5611--5619

\bibitem[{Dr\u{a}gulescu and Yakovenko(2000)}]{DY_2000}
Dr\u{a}gulescu A, Yakovenko V (2000) Statistical mechanics of money. Eur Phys J
  B 17:723--729

\bibitem[{Dr\u{a}gulescu and Yakovenko(2001{\natexlab{a}})}]{DY_2001b}
Dr\u{a}gulescu A, Yakovenko V (2001{\natexlab{a}}) Evidence for the exponential
  distribution of income in the {USA}. Eur Phys J B 20:585--589

\bibitem[{Dr\u{a}gulescu and Yakovenko(2001{\natexlab{b}})}]{DY_2001a}
Dr\u{a}gulescu A, Yakovenko V (2001{\natexlab{b}}) Exponential and power-law
  probability distributions of wealth and income in the {U}nited {K}ingdom and
  the {U}nited {S}tates. Physica A 299:213--221

\bibitem[{Eurostat(2005)}]{EURO_2005}
Eurostat (2005) EUSILC UDB 2005 -- version 5 of August 2009

\bibitem[{Eurostat(2006)}]{EURO_2006}
Eurostat (2006) EUSILC UDB 2006 -- version 4 of March 2010

\bibitem[{Eurostat(2007)}]{EURO_2007}
Eurostat (2007) EUSILC UDB 2007 -- version 3 of March 2010

\bibitem[{Eurostat(2008)}]{EURO_2008}
Eurostat (2008) EUSILC UDB 2008 -- version 2 of August 2010

\bibitem[{Eurostat(2009)}]{EURO_2009}
Eurostat (2009) EUSILC UDB 2009 -- version 3 of March 2012

\bibitem[{Eurostat(2010)}]{EURO_2010}
Eurostat (2010) EUSILC UDB 2010 -- version 1 of March 2012

\bibitem[{Eurostat(2013)}]{EURO}
Eurostat (2013) Statistics on income, social inclusion and living conditions.
  Eurostat, \small
  http://epp.eurostat.ec.europa.eu/portal/page/portal/\\income\_social\_inclusion\_living\_conditions/introduction
  \normalsize

\bibitem[{Ferrero(2005)}]{F_2005}
Ferrero J (2005) The monomodal, polymodal, equilibrium and nonequilibrium
  distribution of money. In: Chatterjee A, Chakrabarti B, Yarlagadda S (eds)
  Econophysics of Wealth Distributions, Springer, Italy, pp 159--167

\bibitem[{Ferrero(2004)}]{F_2004}
Ferrero JC (2004) The statistical distribution of money and the rate of money
  transference. Physica A 341:575--585

\bibitem[{Fischer and Braun(2003)}]{FB_2003}
Fischer R, Braun D (2003) Transfer potentials shape and equilibrate monetary
  systems. Physica A 321:605--618

\bibitem[{Forbes(2013)}]{FORBES}
Forbes (2013) The world's billionaires. Forbes,
  http://www.forbes.com/billionaires/

\bibitem[{Fujiwara et~al(2003)Fujiwara, Souma, Aoyama, Kaizoji, and
  Aoki}]{FSAKA_2003}
Fujiwara Y, Souma W, Aoyama H, Kaizoji T, Aoki M (2003) Growth and fluctuations
  of personal income. Physica A 321:598--604

\bibitem[{Galam(2012)}]{G_2012}
Galam S (2012) Sociophysics -- A Physicist's Modeling of Psycho-political
  Phenomena. Springer, New York

\bibitem[{Gibrat(1931)}]{G_1931}
Gibrat R (1931) Les in{\'e}galit{\'e}s {\'e}conomiques: applications aux
  in{\'e}galit{\'e}s des richesses, {\`a}~la concentration des
  entreprises...d'une loi nouvelle, la loi de l'effet proportionnel. Recueil
  Sirey, Paris

\bibitem[{Haneberg(2004)}]{Han_2004}
Haneberg W (2004) Computational geosciences with Mathematica. Springer, Berlin

\bibitem[{Huang(2004)}]{H_2004}
Huang D (2004) Wealth accumulation with random redistribution. Phys Rev E
  69:057{}103

\bibitem[{Ispolatov et~al(1998)Ispolatov, Krapivsky, and Redner}]{IKR_1998}
Ispolatov S, Krapivsky P, Redner S (1998) Wealth distributions in asset
  exchange models. Eur Phys J B 2:267--276

\bibitem[{Jagielski(2009)}]{mgr}
Jagielski M (2009) Badanie zamo\.{z}no\'{s}ci gospodarstw domowych w {P}olsce
  metodami egzotycznej i tradycyjnej fizyki statystycznej. Master's thesis,
  Faculty of Physics, University of Warsaw, Warsaw, available at Library of
  Institute of Experimental Physics

\bibitem[{Jagielski and Kutner(2013{\natexlab{a}})}]{JK_2013a}
Jagielski M, Kutner R (2013{\natexlab{a}}) Ab initio analysis of all income
  society classes in the {E}uropean {U}nion. Acta Phys Pol A 123:538--541

\bibitem[{Jagielski and Kutner(2013{\natexlab{b}})}]{JK_2013b}
Jagielski M, Kutner R (2013{\natexlab{b}}) Modelling of income distribution in
  the {E}uropean {U}nion with the {F}okker--{P}lanck equation. Physica A
  392:2130--2138

\bibitem[{Kalecki(1945)}]{K_1945}
Kalecki M (1945) On the {G}ibrat distribution. Econometrica 13:161--170

\bibitem[{van Kampen(2011)}]{vanK_1990}
van Kampen N (2011) Stochastic Processes in Physics and Chemistry. Elsevier,
  Amsterdam

\bibitem[{Kim and Yoon(2004)}]{KY_2004}
Kim K, Yoon S (2004) Power law distributions in {K}orean household incomes,
  unpublished paper, available at: http://arxiv.org/abs/cond-mat/0403161

\bibitem[{Kiyotaki and Wright(1993)}]{KW_1993}
Kiyotaki N, Wright R (1993) A search--theoretic approach to monetary economics.
  Am Econ Rev 83:63--77

\bibitem[{Levy and Solomon(1997)}]{LS_1997}
Levy M, Solomon S (1997) New evidence for the power-law distribution of wealth.
  Physica A 242:90--94

\bibitem[{{\L}ukasiewicz and Or{\l}owski(2004)}]{LO_2004}
{\L}ukasiewicz P, Or{\l}owski A (2004) Probabilistic models of income
  distributions. Physica A 344:146--151

\bibitem[{Mandelbrot(1960)}]{M_1960}
Mandelbrot B (1960) The {P}areto-{L}{\'e}y law and the distribution of income.
  Int Econ Rev 1:79--106

\bibitem[{Mandelbrot(1963)}]{M_1963}
Mandelbrot B (1963) New methods in statistical economics. J Pol Econ
  71:421--440

\bibitem[{Matteo et~al(2004)Matteo, Aste, and Hyde}]{MAH_2004}
Matteo TD, Aste T, Hyde S (2004) Exchanges in complex networks: income and
  wealth distributions. In: Mallamace F, Stanley H (eds) The Physics of Complex
  Systems (New Advances and Perspectives), IOS Press, Amsterdam, pp 435--442

\bibitem[{Molico(2006)}]{M_2006}
Molico M (2006) The distribution of money and prices in search equilibrium. Int
  Econ Rev 47:701--722

\bibitem[{Nirei and Souma(2007)}]{NS_2007}
Nirei M, Souma W (2007) A two factor model of income distribution dynamics. Rev
  of Income and Wealth 53:440--459

\bibitem[{Pareto(1897)}]{P_1897}
Pareto V (1897) Cours d'{\'e}conomie politique. L'Universit{\'e} de Lausanne

\bibitem[{Patriarca et~al(2004{\natexlab{a}})Patriarca, Chakraborti, and
  Kaski}]{PCK_2004a}
Patriarca M, Chakraborti A, Kaski K (2004{\natexlab{a}}) Gibbs versus
  non-{G}ibbs distributions in money dynamics. Physica A 340:334--339

\bibitem[{Patriarca et~al(2004{\natexlab{b}})Patriarca, Chakraborti, and
  Kaski}]{PCK_2004}
Patriarca M, Chakraborti A, Kaski K (2004{\natexlab{b}}) Statistical model with
  a standard $\gamma$ distribution. Phys Rev E 70:016{}104

\bibitem[{Patriarca et~al(2005)Patriarca, Chakraborti, Kaski, and
  Germano}]{PCKG_2005}
Patriarca M, Chakraborti A, Kaski K, Germano G (2005) Kinetic theory models for
  the distribution of wealth: power law from overlap of exponentials. In:
  Chatterjee A, Chakrabarti B, Yarlagadda S (eds) Econophysics of Wealth
  Distributions, Springer, Italy, pp 93--110

\bibitem[{Rawlings et~al(2004)Rawlings, Reguera, and Reiss}]{RRR_2004}
Rawlings P, Reguera D, Reiss H (2004) Entropic basis of the {P}areto law.
  Physica A 343:643--652

\bibitem[{Reed(2003)}]{R_2003}
Reed W (2003) The {P}areto law of incomes -- an explanation and an extension.
  Physica A 319:469--486

\bibitem[{Repetowicz et~al(2005)Repetowicz, Hutzler, and Richmond}]{RHR_2005}
Repetowicz P, Hutzler S, Richmond P (2005) Dynamics of money and income
  distributions. Physica A 356:641--654

\bibitem[{Richmond and Solomon(2001)}]{RS_2001}
Richmond P, Solomon S (2001) Power laws are disguised {B}oltzmann laws. Int J
  Mod Phys C 12:333--343

\bibitem[{Richmond et~al(2006)Richmond, Hutzler, Coelho, and
  Repetowicz}]{RHCR_2006}
Richmond P, Hutzler S, Coelho R, Repetowicz P (2006) A review of empirical
  studies and models of income distributions in society. In: Chakrabarti B,
  A~Chakraborti AC, Editors (eds) Econophysics \& Sociophysics: Trends \&
  Perspectives, WILEY-VCH, Weinheim, pp 129--158

\bibitem[{Roehner(2002)}]{R_2002}
Roehner B (2002) Patterns of Speculation. A Study in Observational
  Econophysics. Cambridge University Press, Cambridge

\bibitem[{Scafetta et~al(2004{\natexlab{a}})Scafetta, Picozzi, and
  West}]{SPW_2004a}
Scafetta N, Picozzi S, West B (2004{\natexlab{a}}) An out-of-equilibrium model
  of the distributions of wealth. Quant Finance 4:353--364

\bibitem[{Scafetta et~al(2004{\natexlab{b}})Scafetta, Picozzi, and
  West}]{SPW_2004b}
Scafetta N, Picozzi S, West B (2004{\natexlab{b}}) A trade-investment model for
  distribution of wealth. Physica D 193:338--352

\bibitem[{Silva and Yakovenko(2005)}]{SY_2005}
Silva A, Yakovenko V (2005) Temporal evolution of the ''thermal'' and
  ''superthermal'' income classes in the {USA} during 1983--2001. Europhys Lett
  69:304--310

\bibitem[{Sinha(2006)}]{S_2006}
Sinha S (2006) Evidence for power-law tail of the wealth distribution in
  {I}ndia. Physica A 359:555--562

\bibitem[{Solomon and Richmond(2001)}]{SR_2001}
Solomon S, Richmond P (2001) Power laws of wealth, market order volumes and
  market returns. Physica A 299:188--197

\bibitem[{Solomon and Richmond(2002)}]{SR_2002}
Solomon S, Richmond P (2002) Stable power laws in variable economies;
  {L}otka-{V}olterra implies {P}areto-{Z}ipf. Eur Phys J B 27:257--261

\bibitem[{Souma(2001)}]{S_2001}
Souma W (2001) Universal structure of the personal income distribution.
  Fractals 9:463--470

\bibitem[{Souma and Nirei(2005)}]{SN_2005}
Souma W, Nirei M (2005) Empirical study and model of personal income. In:
  Chatterjee A, Chakrabarti B, Yarlagadda S (eds) Econophysics of Wealth
  Distributions, Springer, Italy, pp 34--42

\bibitem[{Stanley et~al(1996)Stanley, Afanasyev, Amaral, Buldyrev, Goldberger,
  Havlin, Leschhorn, Maass, Mantegna, Peng, Prince, Salinger, Stanley, and
  Viswanathan}]{SAABGHLMMPPSSV_1996}
Stanley H, Afanasyev V, Amaral L, Buldyrev S, Goldberger A, Havlin S, Leschhorn
  H, Maass P, Mantegna R, Peng CK, Prince P, Salinger M, Stanley M, Viswanathan
  G (1996) \textit{ Anomalous fluctuations in the dynamics of complex systems:
  from DNA and physiology to econophysics}. Physica A 224:302--321

\bibitem[{Sutton(1997)}]{S_1997}
Sutton J (1997) Gibrat's legacy. J Econ Lit 35:40--59

\bibitem[{Xi et~al(2005)Xi, Ding, and Wang}]{XDW_2005}
Xi N, Ding N, Wang Y (2005) How required reserve ratio affects distribution and
  velocity of money. Physica A 357:543--555

\bibitem[{Yakovenko and Rosser(2009)}]{YR_2009}
Yakovenko V, Rosser J (2009) Colloquium: Statistical mechanics of money,
  wealth, and income. Rev Mod Phys 81:1703--1725

\bibitem[{Yakovenko and Silva(2005)}]{YS_2005}
Yakovenko V, Silva A (2005) Two-class structure of income distribution in the
  {USA}: Exponential bulk and power-law tail. In: Chatterjee A, Chakrabarti B,
  Yarlagadda S (eds) Econophysics of Wealth Distributions, Springer, Italy, pp
  15--23

\end{thebibliography}

\end{document}